%% file: itwist20iti.tex
\documentclass[conference]{IEEEtran}
\IEEEoverridecommandlockouts

\input{headings_herzet.tex}

\input{def_macro.tex}

\usepackage{cite}
\usepackage{amsmath,amssymb,amsfonts}
\usepackage{algorithmic}
\usepackage{graphicx}
\usepackage{textcomp}
\usepackage{xcolor}
\begin{document}

\title{Translation-invariant interpolation of parametric dictionaries
\thanks{The authors thank the project ``ANR BECOSE'' (ANR-15-CE23-0021) for its support.}
}

\author{\IEEEauthorblockN{Fr\'ed\'eric Champagnat}
\IEEEauthorblockA{\textit{ONERA, The French Aerospace Lab} \\
Palaiseau, France \\
frederic.champagnat@onera.fr}
\and
\IEEEauthorblockN{C\'edric Herzet}
\IEEEauthorblockA{\textit{INRIA Rennes-Bretagne Atlantique} \\
Rennes, France \\
cedric.herzet@inria.fr}
}

\maketitle

\begin{abstract}
In this communication, we address the problem of approximating the atoms of a parametric dictionary $\dico=\{\atom(\param):\param\in\paramSet\}$, commonly encountered in the context of sparse representations in ``continuous'' dictionaries. 
We focus on the case of translation-invariant dictionaries, where the inner product between $\atom(\param)$ and $\atom(\param')$ only depends on the difference $\param-\param'$.
We investigate the following general question: is there some low-rank approximation of $\dico$ which interpolates a subset of atoms $\{\atom(\param_\idxp)\}_{\idxp=1}^\np$ while preserving the translation-invariant nature of the original dictionary?  
We derive necessary and sufficient conditions characterizing the existence of such an ``interpolating'' and ``translation-invariant'' low-rank approximation. 
Moreover, we provide closed-form expressions of such a dictionary when it exists. 
We illustrate the applicability of our results in the case of a two-dimensional isotropic Gaussian dictionary. We show that, in this particular setup, the proposed approximation framework outperforms standard Taylor approximation. \\
\end{abstract}


\section{Introduction}

Sparse representations aim at representing a signal of interest as the linear combination of a few elements of a dictionary~$\dico$. 
Recently, this problem has been reformulated in a ``continuous'' setting, where the elements of $\dico$ are continuously indexed by some parameter $\param$: 
\begin{align}\label{eq:DefDico}
\dico = \kbrace{\atom(\param) \in {\Hs}: \param\in\paramSet}
\end{align}
where $\paramSet$ is a square 
interval of $\Rbb^{\dimParam}$, $\Hs$ is a Hilbert space over the real field $\Rbb$ with inner product $\scalprod{\cdot}{\cdot}$ and induced norm $\normH{\cdot}$, and $\atom(\cdot)$ some continuous function from  $\paramSet$ to $\Hs$, see \eg \cite{Bredies:2013lq,Candes2014Towards,Denoyelle:2018yq}. We adopt the common hypothesis that $\normH{\atom(\param)}=1$ $\forall\param\in\paramSet$. We focus on families of translation invariant parametric dictionaries: we consider the case where \vspace{0.1cm}
\begin{align}\label{eq:translation_invariance}
\scalprod{\atom(\param)}{\atom(\param')}= \scalprod{\atom(\param-\boldsymbol{\tau})}{\atom(\param'-\boldsymbol{\tau})}, \vspace{0.1cm}
\end{align}
that is the inner product between two atoms of the dictionary only depends on $\param-\param'$. 
This setup is ubiquitous in many physical, chemical or biological problems where the observed signal is the linear combination of shifted copies of the system's impulse response, see \eg~\cite{Denoyelle:2018yq,Yang:2018kk}.

Continuous dictionaries contain an infinite uncountable number of elements and induce therefore new difficulties. Trivial operations in the discrete setting may become challenging in the continuous framework. For instance, consider the well-known ``atom selection'' problem \cite{Pati_asilomar93,Jaggi_2013}:
\begin{align}\label{eq:atom_selection_step}
\mathrm{Find}\ \param^\star &= \kargmax_{\param\in\paramSet} \scalprod{\atom(\param)}{\rr} \quad \mbox{for some $\rr\in\Hs$},
\end{align}
which only involves the evaluation of a finite number of inner products in the discrete setting but can turn out to be a difficult optimization task in the continuous framework. 

In order to circumvent this problem, several contributions of the literature have proposed to tackle the infinite-dimensional nature of continuous dictionaries by resorting to ``low-rank'' approximations of $\atom(\param)$. 
More formally, the idea consists in approximating the elements of $\dico$ as linear combinations of a few vectors $\{\vapprox_\idxv\}_{\idxv=1}^\nv$: 
 \begin{align}\label{eq:lr_atom_approx}
\atomapprox(\param) \triangleq \sum_{\idxv=1}^{\nv} \vapprox_\idxv \coef_\idxv(\param)
\end{align}
for some functions $\{\coef_\idxv:\paramSet\rightarrow \Rbb\}_{\idxv=1}^\nv$. 
This approach was for example used in \cite{Ekanadham2011Recovery} and \cite{Knudson:2014qa} to transform ``continuous'' sparse-representation problems into approximate (but tractable) finite-dimensional ones.~The quality of the approximation obtained with this strategy obviously depends on the choice of the vectors  $\{\vapprox_\idxv\}_{\idxv=1}^\nv$ and functions $\{\coef_\idxv\}_{\idxv=1}^\nv$. 
Several options were considered in \cite{Ekanadham2011Recovery,Knudson:2014qa,Champagnat:2019ly}. In \cite{Ekanadham2011Recovery}, the authors introduced the ``Taylor'' and ``polar'' approximations: the former is based on a Taylor decomposition of $\atom(\param)$; the latter is constructed so that $\atomapprox(\param)$ has a unit norm $\forall\param\in\paramSet$ and interpolates $\atom(\param)$ for some $\{\param_\idxp\}_{\idxp=1}^3$.~In \cite{Knudson:2014qa}, the authors suggested to use a singular-value decomposition of $\atom(\param)$ to identify the approximation subspace minimizing the projection error in a $\normH{\cdot}$-sense.
In \cite{Champagnat:2019ly}, the authors pointed out some connection between the approximations considered in \cite{Ekanadham2011Recovery} and \cite{Knudson:2014qa}.

The present communication is mainly of theoretical nature. We address the following general question: is there some low-rank approximation of the form \eqref{eq:lr_atom_approx} which: \textit{(i)} interpolates the original dictionary over some subset of parameters $\{\param_\idxp\}_{\idxp=1}^{\np}$; \textit{(ii)}~preserves the translation invariance of the original dictionary? 
The formal statement of this question is given below, all the formal derivations are available in \cite{Champagnat:2019wb}.

\section{Problem statement and main result}\label{sec:GPA_Main}
We focus on low-rank approximations of the form \eqref{eq:lr_atom_approx} obeying the two following properties: \vspace{0.1cm}
\begin{itemize}
\item \textit{Interpolation}:
\begin{align}\label{eq:Gen_interpolation_property}
\atomapprox(\param) &= \atom(\param) \quad \mbox{for $\param\in\{\param_\idxp\}_{\idxp=1}^{\np}$}. 
\end{align}
\item \textit{Translation invariance}: 
\begin{align}
\scalprod{\atomapprox(\param)}{\atomapprox(\param')}&= \lambda_0 + \sum_{\idxk=1}^\nk \lambda_\idxk \cos(\ktranspose{\wc}_\idxk(\param-\param'))\label{eq:Tinv_LR_Kernel1MainB}
\end{align}
for some $\{\lambda_\idxk\}_{\idxk=0}^\nk$ and $\{\wc_\idxk\}_{\idxk=1}^\nk$. \vspace{0.2cm}
\end{itemize}
\noindent
Property~\eqref{eq:Gen_interpolation_property} imposes that the approximation $\atomapprox(\param)$ perfectly interpolates the original atom $\atom(\param)$ for some values of the parameters $\{\param_j\}_{\idxp=1}^{\np}$.~Property \eqref{eq:Tinv_LR_Kernel1MainB} enforces the inner product of the approximated atoms to obey a ``raised-cosine'' kernel.~In particular, we note that satisfying \eqref{eq:Tinv_LR_Kernel1MainB} ensures the translation-invariance of $\atomapprox(\param)$.

In the rest of this section, we answer the two following questions: 1) is there some low-rank approximation $\dicoapprox=\{\atomapprox(\param): \param\in\paramSet\}$ verifying properties \eqref{eq:Gen_interpolation_property} and \eqref{eq:Tinv_LR_Kernel1MainB}? 
2) if such an approximation exists, how to build it? 

Before presenting our main result, we make two important remarks, which will allow us to give a more formal statement of the two above questions.  

First, in order to simplify our exposition, we assume 
 that $\{\lambda_\idxk\}_{\idxk=0}^\nk$ and $\{\wc_\idxk\}_{\idxk=1}^\nk$ are such that the right-hand side of \eqref{eq:Tinv_LR_Kernel1MainB} defines a semidefinite kernel.  This assumption makes sense since the kernel induced by any family of parametric atoms must necessarily be positive semidefinite.
Second, consistency of the different decompositions imply constraints on parameters $(J,K,L)$, 
more precisely 
 \begin{align}
\np &=\nv\label{eq:defJ2k}\\
\nk &=
\kfloor{\nv/2} \label{eq:defL2k}
\end{align}
where $\kfloor{\cdot}$ returns the greatest integer less than or equal to its input argument. 
Moreover $\lambda_0=0$ if $\nv$ is even and $\lambda_0>0$ if $\nv$ is odd.
We also assume that the interpolated vectors $\{\atom(\param_\idxp)\}_{j=1}^{\np}$ are linearly independent. 
The main question addressed herefafter thus takes the following form:
\begin{question}
 Is there a rank-\nv approximation \eqref{eq:lr_atom_approx} that interpolates a family of $\nv$ linearly independent atoms $\{\atom(\param_\idxp)\}_{\idxp=1}^{\nv}$ and verifies \eqref{eq:Tinv_LR_Kernel1MainB} for some 
$\{\lambda_\idxk\}_{\idxk=0}^{\kfloor{\nv/2}}$  and $\{\wc_\idxk\}_{\idxk=1}^{\kfloor{\nv/2}}$? If so, how to build such an approximation? \vspace{0.1cm}
\end{question}

The answer to this question is provided in 
 Theorem~\ref{Theorem:polar_decomposition} below.  
We present necessary and sufficient conditions that ensure the existence of the desired approximation. 
Moreover, when it exists, we provide the closed-form expressions of $\{\vapprox_\idxv \}_{\idxv=1}^\nv$ and $\{\coef_\idxv(\param)\}_{\idxv=1}^\nv$ defining the approximation. \vspace{0.2cm}
 Our main result relates the existence of a low-rank approximation verifying \eqref{eq:Gen_interpolation_property}-\eqref{eq:Tinv_LR_Kernel1MainB} to some constraints on the Gram matrix of the family of interpolated atoms $\{\atom(\param_\idxp)\}_{\idxp=1}^{\nv}$. 
More specifically, letting 
\begin{align}\label{eq:defGram}
\G \triangleq [\scalprod{\atom(\param_\idxpB)}{\atom(\param_\idxp)}]_{\idxpB,\idxp}\in\Rbb^{\nv\times \nv},
\end{align}
the following result holds:
\begin{theorem}
\label{Theorem:polar_decomposition}
Let $\nv\in\kN$. If there exists a family of atoms \eqref{eq:lr_atom_approx} satisfying 
\eqref{eq:Gen_interpolation_property}-\eqref{eq:defL2k} then \vspace{0.1cm}
\begin{equation}
\G(\idxpB,\idxp) = \lambda_0+\sum_{\idxk=1}^\nk \lambda_\idxk \cos(\ktranspose{\wc}_\idxk(\param_\idxpB-\param_\idxp)) \quad \forall \idxpB,\idxp
\label{eq:constrGram}
\end{equation}

for some $\{\lambda_\idxk\}_{\idxk=0}^\nk$, $\{\wc_\idxk\}_{\idxk=1}^\nk$.

Conversely, assume there exist $\{\lambda_\idxk\}_{\idxk=0}^\nk$, $\{\wc_\idxk\}_{\idxk=1}^\nk$ and $\{\param_\idxp\}_{\idxp=1}^{\nv}$ such that \eqref{eq:constrGram}  holds. 
Then, low-rank approximation \eqref{eq:lr_atom_approx} 
with
\begin{align}
\coef_\idxv(\param) &= \lambda_0 + \sum_{\idxk=1}^\nk \lambda_\idxk \cos(\ktranspose{\wc}_{\idxk}(\param-\param_\idxv))\label{eq:polar_dicoapprox_defA}\\
\vapprox_\idxv        &= \sum_{\idxp=1}^{\nv} \atom(\param_{\idxp})\, \G^{-1}({\idxv,\idxp})\label{eq:polar_dicoapprox_defB},
\end{align}
verifies \eqref{eq:Tinv_LR_Kernel1MainB} and interpolates $\{\atom(\param_\idxp)\}_{\idxp=1}^{\nv}$.
\end{theorem}

A proof of this result is available in \cite{Champagnat:2019wb}. 
Theorem~\ref{Theorem:polar_decomposition} provides necessary and sufficient conditions for our ``Main Question" to have a positive answer.~Interestingly, we see that the existence of a low-rank decomposition verifying \eqref{eq:Gen_interpolation_property}-\eqref{eq:Tinv_LR_Kernel1MainB} is exclusively conditioned on the existence of some particular factorization of the Gram matrix $\G$ (see condition \eqref{eq:constrGram}). Hence, Theorem~\ref{Theorem:polar_decomposition} transforms the question of the existence of an interpolating and translation-invariant low-rank approximation into an algebraic problem where one must find a set of parameters $\{\lambda_\idxk\}_{\idxk=0}^\nk$, $\{\wc_\idxk\}_{\idxk=1}^\nk$ verifying equation~\eqref{eq:constrGram}. 

The existence 
of decomposition~\eqref{eq:constrGram} will depend on the nature of the original dictionary $\dico$ and the choice of the interpolation parameters $\{\param_\idxp\}_{\idxp=1}^{\nv}$. Providing a general answer to this algebraic problem is therefore a broad question which is out of the scope of this communication. 
Note however that in the case of interpolated atoms on a regular  grid, the Gram matrix problem decomposition is linked to the well-documented Vandermonde decomposition problem \cite{Yang2016Vandermonde}. The Vandermonde decomposition is not fully resolved in the multi-dimensional case but we provide a solution to this problem in the particular case of a separable isotropic dictionary $\dico$. 

To conclude this section, let us mention that the expressions of the vectors $\{\vapprox_\idxv\}_{\idxv=1}^{\nv}$ and functions $\{\coef_\idxv\}_{\idxv=1}^{\nv}$ provided in Theorem~\ref{Theorem:polar_decomposition} only involve simple operations and are thus easy to evaluate numerically. Hence, the construction a low-rank dictionary verifying \eqref{eq:Gen_interpolation_property}-\eqref{eq:Tinv_LR_Kernel1MainB} is straightforward once the set of paremeters $\{\lambda_\idxk\}_{\idxk=0}^\nk$  and $\{\wc_\idxk\}_{\idxk=1}^\nk$ satisfying conditions \eqref{eq:constrGram} has been identified. \vspace{0.2cm}

We will illustrate the applicability of our results in the case of a two-dimensional isotropic Gaussian dictionary. We show that, in this particular setup, the proposed approximation framework outperforms standard Taylor approximation.

\bibliographystyle{IEEEbib}
\bibliography{MyBibli}

\end{document}

%% file: headings_herzet.tex
\providecommand*{\fullref}[1]{\hyperref[{#1}]{\cref*{#1}. \nameref*{#1}}}
\providecommand*{\Fullref}[1]{\hyperref[{#1}]{\Cref*{#1}. \nameref*{#1}}}

\usepackage{amssymb} 
\usepackage{amsmath} 
\usepackage{amsfonts} 
\usepackage{amsthm}
\usepackage[setbb,setcolon]{kmath} 			
\usepackage[noadjust]{cite}       

\def\0{{\mathbf 0}}
\def\1{{\mathbf 1}}

\def\G{{\mathbf G}}

\def\aa{{\boldsymbol{a}}}

\def\rr{{\boldsymbol r}}

\def\vv{{\boldsymbol v}}

\def\Rbb{{\mathbb R}}

\def\eg{\textit{e.g.},\xspace}

\newkmacro\sign[1][{\cdot}]{\mathrm{sign}\left({#1}\right)} 
\newkmacro\interior[1][{\cdot}]{\mathrm{int}\left({#1}\right)}
\newkmacro\bd[1][{\cdot}]{\mathrm{bd}\left({#1}\right)}
\newkmacro\conv[1][{\cdot}]{\mathrm{conv}\left({#1}\right)}
\newkmacro\card[1][{\cdot}]{\mathrm{card}\left({#1}\right)}
\newkmacro\spa[1][{\cdot}]{\mathrm{span}\left({#1}\right)}
\newkmacro\nul[1][{\cdot}]{\mathrm{null}\left({#1}\right)}
\newkmacro\deter[1][{\cdot}]{\mathrm{det}\left({#1}\right)}
\newkmacro\tr[1][{\cdot}]{\mathrm{tr}\left({#1}\right)}

\newtheorem{theorem}{Theorem}

\newtheorem*{question}{Main Question}


%% file: def_macro.tex
\newcommand{\Hs}{\ensuremath{{\cal{H}}}\xspace}

\newcommand{\dico}{\ensuremath{\mathcal{A}}\xspace}
\newcommand{\atom}{\ensuremath{\aa}\xspace}

\newcommand{\atomapprox}{\ensuremath{\hat{\atom}}\xspace}

\newcommand{\dicoapprox}{\ensuremath{\hat{\dico}}\xspace}

\newcommand{\idxk}{\ensuremath{\ell}\xspace}

\newcommand{\nk}{\ensuremath{L}\xspace}

\newcommand{\paramel}{\ensuremath{\theta}\xspace}
\newcommand{\param}{\ensuremath{\boldsymbol{\paramel}}\xspace}

\newcommand{\paramSet}{\ensuremath{\Theta}\xspace}
\newcommand{\dimParam}{\ensuremath{d}\xspace}

\newcommand{\idxp}{\ensuremath{j}\xspace}
\newcommand{\idxpB}{\ensuremath{i}\xspace}
\newcommand{\np}{\ensuremath{J}\xspace}

\newcommand{\wcel}{\ensuremath{\omega}\xspace}
\newcommand{\wc}{\ensuremath{\boldsymbol{\wcel}}\xspace}

\newcommand{\vapprox}{\ensuremath{\vv}\xspace}
\newcommand{\coef}{\ensuremath{c}\xspace}

\newcommand{\idxv}{\ensuremath{k}\xspace}

\newcommand{\nv}{\ensuremath{K}\xspace}

\newcommand{\scalprod}[2]{\ensuremath{\kangle{#1,#2}\xspace}}

\newcommand{\normH}[1]{\ensuremath{\kvvbar{#1}\xspace}}